\documentclass[11pt]{article}
\usepackage{amsmath}
\usepackage{amssymb, amsfonts}
\usepackage{psfrag,graphicx}
\usepackage{mathrsfs} 
\usepackage{dsfont}
\usepackage{epsfig}
\usepackage{slashed}

\newcommand{\tr}{\,\mathrm{Tr}}

\makeatletter\renewcommand{\section}{\@startsection
{section}{1}{\z@}{-3.5ex plus -1ex minus
    -.2ex}{2.3ex plus .2ex}{\bf }}

\makeatletter\renewcommand{\subsection}{\@startsection{subsection}{2}{\z@}{-3.25ex
plus -1ex minus
   -.2ex}{1.5ex plus .2ex}{\it }}
\makeatletter\renewcommand{\subsubsection}{\@startsection{subsubsection}{3}{-2.45ex}{-3.25ex
plus -1ex minus -.2ex}{1.5ex plus .2ex}{\it }}

\makeatletter \@addtoreset{equation}{section}
\def\be{\begin{equation}}
\def\ee{\end{equation}}
\def\beqa{\begin{eqnarray}}
\def\eeqa{\end{eqnarray}}
\def\nn{\nonumber}

\begin{document}

\begin{titlepage}

\begin{flushright}
{\bf  \today} 
\end{flushright}

\vskip 2cm

\begin{center}
\centerline{\Large \bf Spontaneous Lorentz Violation: The Case of Infrared QED}

\vskip 5em

\centerline{\large \bf A.~P.~Balachandran\footnote{bal@phy.syr.edu} $^{a,d}$, S. K\"{u}rk\c{c}\"{u}o\v{g}lu\footnote{kseckin@metu.edu.tr} $^{b}$, 
A. R. de Queiroz\footnote{amilcarq@unb.br} $^{c,}$ and S. Vaidya\footnote{vaidya@cts.iisc.ernet.in} $^{d}$
}
\vskip 1em

\end{center}

\centerline{\sl $^a$ Physics Department, Syracuse University, Syracuse, NY, 13244-1130, USA. }
\centerline{\sl $^b$ Middle East Technical University, Department of Physics, 06800, Ankara, Turkey.}
\centerline{\sl $^c$ Instituto de F\'{\i}sica, Universidade de Bras\'{\i}lia, Caixa Postal 04455, 70919-970, Bras\'{\i}lia, DF, Brazil. }
\centerline{\sl $^d$Centre for High Energy Physics, Indian Institute of Science, Bangalore 560012, India.}

\vskip 5em

\begin{abstract}

It is by now clear that infrared sector of QED has an intriguingly complex structure. Based on earlier pioneering works on this subject, two of us recently proposed a simple modification of QED by constructing a generalization of the $U(1)$ charge group of QED to the ``Sky'' group  incorporating the known spontaneous Lorentz violation due to infrared photons, but still compatible in particular with locality \cite{Bal-Sachin}. There it was shown that the ``Sky" group is generated by the algebra of angle dependent charges and a study of its superselection sectors has revealed a manifest description of spontaneous breaking of Lorentz symmetry. We further elaborate this approach here and investigate in some detail the properties of charged particles dressed by the infrared photons. We find that Lorentz violation due to soft photons may be manifestly codified in an angle dependent fermion mass modifying therefore the fermion dispersion relations. The fact that the masses of the charged particles are not Lorentz invariant affects their spin content too.Time dilation formulae for decays should also get corrections. We speculate that these effects could be measured possibly in muon decay experiments. 

\end{abstract}

\end{titlepage}

\section{Introduction}

In quantum field theory (QFT), observables are local and generate the algebra of local observables $\mathscr{A}$. In contrast, time evolution and global symmetries are not local. In Lagrangian field theories, their generators involve integrals of fields over all of space. They are thus not elements of $\mathscr{A}$, but instead generate automorphisms of $\mathscr{A}$. They are elements of the automorphism group ${\rm Aut}\mathscr{A}$ of $\mathscr{A}$.

The group of automorphisms of $\mathscr{A}$ generated by conjugation using unitary elements of $\mathscr{A}$ is the group ${\rm Inn}\mathscr{A}$ of inner automorphisms. It is a normal subgroup of ${\rm Aut}\mathscr{A}$. The group ${\rm Aut}\mathscr{A}/{\rm Inn}\mathscr{A}$ is the outer automorphism group ${\rm Out}\mathscr{A}$.  

The automorphisms generated by some global symmetries may be equivalent to inner ones. If that is not the case, then they define non-trivial elements of  ${\rm Out}\mathscr{A}$.

In QFT, we choose an irreducible representation $\rho$ or a superselection sector of $\mathscr{A}$. It can happen that a global symmetry transformation cannot be implemented by a unitary or antiunitary operator in the representation space $\mathscr{H}$ of $\mathscr{A}$. In that case, the symmetry is said to be spontaneously broken. Since elements of $\mathscr{A}$ act by definition on $\mathscr{H}$,  ${\rm Inn}\mathscr{A}$ cannot be spontaneously broken. So that can happen only if $S\notin {\rm Inn}\mathscr{A}$.  

Spontaneous breaking by a Higgs field can be understood in this framework. Thus no local operation can change the asymptotic expression $\phi_\infty$ of the Higgs field $\phi$. Hence $\phi_\infty$ is a label for the representation $\rho$. If a symmetry changes $\phi_\infty$, it changes the representation. Hence it is spontaneously broken.

The mechanism of spontaneous breaking can be illustrated even with the $3\times 3$ matrix algebra $M_3(\mathbb{C})$. In its irreducible representation $\rho$, which is three-dimensional, its unitary subgroup $U_3$ is represented irreducibly by a representation we can name as $\underline{3}$. Complex conjugation is an anti-linear automorphism of $M_3(\mathbb{C})$. It changes $\underline{3}$ to the inequivalent representation $\underline{\bar{3}}$ and hence $\rho$ to the inequivalent representation $\bar{\rho}$. This anti-linear automorphism is thus spontaneously broken in the representation $\rho$.

The Poincar\'e  group is an automorphism group of the observables of QED. It transforms elements of $\mathscr{A}$ nontrivially. Instead, the electric charge $Q$, at least classically, is the total electric flux at infinity and hence commutes with all elements of $\mathscr{A}$. Its different values $q$ go into the labels for the different irreducible representations of $\mathscr{A}$. They are similar to the Casimir operators of Lie algebras. Since $Q$ is Poincar\'e invariant, Poincar\'e symmetry is compatible with charge superselection. The latter does not cause spontaneous Poincar\'e violation.

In QED, infrared photons accumulate at infinity and create a non-zero electromagnetic field $f_{\mu \nu}$ there as nicely shown by Buchholz \cite{Buchholz2} \cite{Buchholz1} and Fr\"ohlich et al.\cite{Frohlich197961, Frohlich1979241}. Since local operations cannot affect $f_{\mu \nu}$, this field also labels superselection sectors. But $f_{\mu \nu}$ is not Lorentz invariant. Therefore Lorentz symmetry is spontaneously broken in QED. 

The group $U(1)$ of QED is based on electric charge which classically is a measure of the net flux of electric field $\vec{E}$ ($E_{i}=f_{0i}$) on the sphere $S^2_\infty$ at infinity. In previous work, we extended $U(1)$ to the ``Sky'' group $\mathscr{G}_{\rm sky}$ which is sensitive to all the partial waves of ${\vec E}$ on $S^2_\infty$. It is superselected because $f_{0i}$ is. But elements of $\mathscr{G}_{\rm sky}/U(1)$ are not Lorentz invariant and cause spontaneous Lorentz violation.

In quantum theory, the infrared cloud on $S^2_\infty$ is incorporated in the state vectors, the charged states being dressed by a coherent state of photons \cite{Eriksson, Gervais, Frohlich197961, Frohlich1979241}. We have constructed the dressing operator in \cite{Bal-Sachin} using a closed form $\omega$. Previous work \cite{Frohlich197961, Frohlich1979241} determine the coherent state of the photon and hence $\omega$. Thus we can write the dressed charged particle states. A simple twist of the electron (or charged particle) mass is sensitive to the coherent state. The twisted mass is not Lorentz invariant. It affects the dispersion relation, the spin content of the particle and time dilatations in decays and life times. These can be measured. Naturally it also leads to Lorentz non-invariant scattering amplitudes in charged sectors which too can be measured.

In section 2, we recall the vertex operator dressing the charged vector states with a cloud of infrared photons. It  is gauge, but not in general Lorentz invariant. A new result we describe is its incorporation in the Lagrangian. We can set  it equal to $1$ by adding the term proportional to 
\begin{equation}
\int F\wedge \omega, \qquad F : \textrm{Electromagnetic field} \,, \qquad \omega : \textrm{closed two-form} \,,
\label{omegaterm}
\end{equation}
to the action. It is remarkably close to the $\theta$-term
\begin{equation}
\frac{\theta}{8\pi^2}\int \tr~F\wedge F, \quad F : \textrm{curvature two-form} \,,
\label{thetaterm}
\end{equation}
in QCD which induces $P$ and $T$ violation. Just like the $\theta$-term, \eqref{omegaterm} is a surface term and does not affect equations of motion.

By QED, we always mean QED in $(3+1)$-dimensional spacetime, or QED$_4$. But we can also consider QED$_3$ in $(2+1)$-dimensional spacetime where analogous considerations regarding infrared photons apply \cite{Balachandran:2012md}. In that case, the dressing operator involves a closed one-form $\omega$, so that on $\mathbb{R}^{2,1}$, $\omega=d\alpha$, with $\alpha$ a scalar function. The term in the action which absorbs the vertex operator from state vectors is proportional to 
\begin{equation}
\int F\wedge d\alpha
\end{equation}
which is close to the standard Chern-Simons term. In order to see this, we can interpret $A$ in $\int A \wedge F$ as $d\alpha$.

Section 3 reviews the proposed mass twist introduced in \cite{Bal-Sachin}. We establish that the twist is compatible with locality. Still it is affected by the infrared cloud at $S^2_\infty$. Thus inertia is affected by the ``vacuum'' as in Mach's principle. In a similar manner, the Higgs field at infinity induces vector meson masses. Section 4 briefly indicates the changes in our considerations for QED$_3$. In particular, in this case the $\omega$ in the vertex operator is a closed one-form, $\omega = d\alpha$ on $\mathbb{R}^{2,1}$. Section 5 summarizes the part of Fr\"ohlich et al. \cite{Frohlich1979241} helpful in the construction of $\omega$. In section 6, we examine the modified dispersion relations of charged particles which violate Lorentz invariance. It seems possible to measure this violation. There are also similar violations in scattering amplitudes which are also in principal measurable.

For related recent work with emphasis on BMS group, see \cite{Strominger:2013lka, Strominger:2013jfa, He:2014laa, Cachazo:2014fwa}.

Extensive work on Lorentz violation using effective Lagrangians has been done by Kostelecky and collaborators \cite{Colladay:1998fq,Kostelecky:2009zp,Kostelecky:2013rta}. A study of the relation of their approach and ours is yet to be done. Up-to-date experimental constraints on Lorentz violation may be reached from the data tables in \cite{Kostelecky:2008ts}. An easily accessible review on both terrestrial and astrophysical constraints on Lorentz violation may be found in \cite{Mattingly}.

\section{The Vertex Operator for QED$_4$ Coherent States}

\subsection{Preliminaries}

The gauge transformations generated by the Gauss law will be denoted as the group $\mathcal{G}^\infty_0$, the subscript $0$ denoting that it is connected to identity. Its elements $g$ approach the identity when its spatial argument $\vec{x}$ approaches infinity,
\begin{equation}
  \mathcal{G}^\infty_0=\left\langle g: \mathbb{R}^3\to U(1): \qquad g(\vec{x}) \to e \quad \textrm{ as } \quad |\vec{x}|\to\infty   \right\rangle.
\end{equation}
Its Lie algebra is spanned by
\begin{equation}
  G(\Lambda)=\int d^3x~\left(-E_i\partial_i\Lambda + \Lambda J_0 \right)(\vec{x}),
\end{equation}
where $E_i$ is the electric field, $J_0$ is the charge density and
\begin{equation}
 \Lambda(\vec{x})\to 0 \quad \textrm{ as } \quad |\vec{x}|\to \infty, \quad \textrm{i.e. } \Lambda\in\mathcal{C}_0^\infty,
\end{equation}
where the subscript $0$ indicates compact support and superscript $\infty$ indicates infinite differentiability, both of which are standard notations. The Gauss law is imposed in the theory by requiring that $G(\Lambda)$ vanishes for any choice of the test function $\Lambda\in\mathcal{C}_0^\infty$ so that ${\mathcal G}^\infty_0 \to \{\mathds{1}\}$ on quantum states.

Another manner to state this requirement is the following. If $\mathscr{C}$ is the space of connections and $\mathcal{G}^\infty_0$ acts on $\mathscr{C}$ by gauge transformations, the principle bundle for gauge theories without matter is $\mathscr{C}/\mathcal{G}^\infty_0$, that is, the quotient of $\mathscr{C}$ by the action of $\mathcal{G}^\infty_0$.

The generators $Q_0(\chi)$ of the charge group also come from the Gauss law,
\begin{equation}
  Q_0(\chi)=\int d^3x~\left(-E_i\partial_i\chi + \chi J_0 \right)(\vec{x}),
\end{equation}
but now $\chi$ is required to approach a constant at infinity,
\begin{equation}
  \chi(\vec{x})\to \chi_\infty \quad \textrm{ as } \quad |\vec{x}|\to\infty.
\end{equation}
The charge group they generate will be denoted by $\mathcal{G}$. The group $\mathcal{G}^\infty_0$ is normal in $\mathcal{G}$. Since ${\mathcal G}^\infty_0 \to \{\mathds{1}\} $ on quantum states, the effective group is $\mathcal{G}/\mathcal{G}^\infty_0\approx U(1)$. The normalized charge is jus
\begin{equation}
  Q_0~ := ~Q_0(\chi)|_{\chi_\infty=1}.
\end{equation}

For QED with spatial slice $\mathbb{R}^3$, we can choose $\chi(\vec{x})=1$ for all $\vec{x}$ conveniently since $G(\Lambda)\to 0$ in quantum theory. That gives the standard expression
\begin{equation}
  Q_0=\int d^3x J_0
\end{equation}
for charge.

The Sky group $\mathcal{G}_{\rm sky}$ has generators 
\begin{align}
  Q(\chi)&=\int d^3x \left(-E_i\partial_i \chi +\chi J_0 \right)(\vec{x}), \\
  \chi(\vec{x}) &\equiv \chi(r\hat{n}) \stackrel{r\to\infty}{\longrightarrow} \chi^\infty(\hat{n}) \,, \quad r = |\vec{x}| \,, \quad \hat{n} = \frac{\vec{x}}{|\vec{x}|}
\end{align}
where the function $\chi^\infty$ on $S^2_\infty$ need not be constant. As a result, $Q(\chi)$ need not even be rotationally invariant modulo a $G(\Lambda)$. Hence $\mathcal{G}_{\rm sky}$ breaks Lorentz invariance.

The Sky group acts trivially on the charge zero sector which has no infrared cloud. But in charged sectors, state vectors are twisted by a coherent state vector of the infrared cloud, and hence, $\mathcal{G}_{\rm sky}$ can act by a non-trivial representation breaking Lorentz invariance.

We note that $\mathcal{G}_{\rm sky}$ has one-dimensional irreducible representations. That is because it is abelian. In fact, since the action of $Q(\chi)$ on quantum state vectors depends only on $\chi^\infty({\hat n})$, we can write,
\begin{equation}
  Q(\chi)=\sum_{lm} Q(\chi^\infty_{lm}Y_{lm}),
\end{equation}
where 
\begin{equation}
  \chi^\infty(\hat{n})=\sum_{lm}\chi^\infty_{lm}Y_{lm}(\hat{n}).
\end{equation}
All the $Q(\chi^\infty_{lm}Y_{lm})$ commute among themselves. Hence each of them generates a one-dimensional abelian group.

Suppose we are given a standard representation of $\mathcal{G}_{\rm sky}$,
\begin{equation}
  Q(\chi)|\cdot\rangle=\chi^\infty_{00}Q_0|\cdot\rangle.
\end{equation}
Then the vertex operator
\begin{equation}
  V(\omega)=e^{i \int A\wedge \omega},
\end{equation}
where $\omega$ is a closed two-form,
\begin{equation}
  d\omega=0,
\end{equation}
maps $|\cdot\rangle$ to the infrared-dressed states $|\cdot\rangle_\omega$ as we show below:
\begin{align}
  |\cdot\rangle_\omega &= e^{i \int A\wedge \omega}|\cdot\rangle_0, \label{IRdressed} \\
  |\cdot\rangle_0 &\equiv |\cdot\rangle.
\end{align}
But first note that $V(\omega)$ is ``gauge invariant'', that is, that it commutes with $\mathcal{G}^\infty_0$,
\begin{equation}
  \left[G(\Lambda),\int A\wedge \omega \right] = \int d\Lambda \wedge \omega = -\int \Lambda  d\omega=0,
\end{equation}
since $\Lambda|_{S^2_\infty}=0$.

Hence $V(\omega)|\cdot\rangle$ is a Gauss law compatible vector (assuming that is the case with $|\cdot\rangle$).

But $Q(\chi)$ does not commute with $V(\omega)$. Let $\omega^{\infty}$ be the asymptotic expression for $\omega$,
\begin{equation}
  \omega^\infty(\hat{n})=\lim_{r\to\infty} \omega(r\hat{n}),
\end{equation}
while let $\chi^\infty$ be the asymptotic $\chi$ as before,
\begin{equation}
  \chi^\infty(\hat{n})=\lim_{r\to\infty} \chi(r\hat{n}).
\end{equation}
Then with
\begin{equation}
  U(\chi)=e^{iQ(\chi)},
\end{equation}
we obtain
\begin{equation}
  U(\chi)V(\omega)=c(\chi^\infty,\omega^\infty) V(\omega) U(\chi),
\end{equation}
with the central element
\begin{equation}
  c(\chi^\infty,\omega^\infty)=\exp\left(-i\int_{S^2_\infty} \omega^\infty\chi^\infty  \right).
\end{equation}
These $c$'s generate the centre of the algebra of $U$'s and $V$'s. This algebra resembles the Weyl algebra.

If 
\begin{equation}
  U(\chi)|\cdot\rangle=e^{i\chi^\infty_{00}Q_0}|\cdot\rangle,
\end{equation}
where $|\cdot\rangle$ denotes any vector transforming just by $U(1)$ of charge of Sky, then
\begin{equation}
  V(\omega)|\cdot\rangle=|\cdot\rangle_\omega
\end{equation}
can transform non-trivially under Sky,
\begin{equation}
  U(\chi)|\cdot\rangle_\omega=c(\chi^\infty,\omega^\infty)|\cdot\rangle_\omega.
\end{equation}
If $\omega^\infty$ is not zero, then the sector $\mathscr{H}_\omega$ of the Hilbert space spanned by $\{ |\cdot\rangle_\omega\}$ breaks Lorentz symmetry.

\subsection{Incorporation of Vertex Operator in Lagrangian}

Let $E_j$ denote the electric field. It is conjugate to the potential $A_j$:
\begin{equation}
\left[A_i(\vec{x}),E_j(\vec{y}) \right]=i\delta_{ij}\delta^3(\vec{x}-\vec{y} \,) 
\end{equation}
at equal times. Now, let us write $\omega = \omega_{kl} dx^k\wedge dx^l$,
\begin{align}
_\omega\langle\cdot| E_j |\cdot'\rangle_\omega &={}_0\langle\cdot| V(\omega)^{-1}E_j V(\omega) |\cdot'\rangle_0 \nonumber \\&= {}_0\langle\cdot| E_j+\left[(-iQ_0)\int \ast \, \omega_i A_i~,~E_j \right] |\cdot'\rangle_0 \nonumber \\ 
&= {}_0\langle\cdot| E_j + \ast \, \omega_j |\cdot'\rangle_0 \,,
\label{electricfield1}
\end{align}
where $\ast \omega_j = \epsilon_{jkl} \omega_{kl}$. Note that in \eqref{electricfield1}, the ket $|\cdot'\rangle$ and bra $\langle \cdot|$ may represent different vectors. Thus the Hamiltonian of $F_{\mu\nu}$,
\begin{equation}
H=\frac{1}{2}\int d^3x \left(\vec{E}^2+\vec{B}^2 \right), \quad \textrm{ with } \quad \vec{B} : \textrm{magnetic field},
\end{equation}
fulfils
\begin{align}
  _\omega\langle\cdot| H |\cdot'\rangle_\omega & = {}_0 \langle\cdot| V(\omega)^{-1} H V(\omega) |\cdot'\rangle_0
 =: {}_0\langle\cdot| \widehat{H} |\cdot'\rangle_0, \\
  \widehat{H} &= \frac{1}{2}\int d^3x \left(\vec{\widehat{E}}^2+\vec{B}^2 \right) \,, \quad \widehat{E}_i = E_i + \ast \, \omega_i.
\end{align}
The field conjugate to $A_i$ is thus shifted from $E_i$ to $\widehat{E}_i$ for the zero-twist state vectors . We can accomplish this shift by adding to the QED Lagrangian density $\mathscr{L}$ a term $\Delta {\mathscr{L}}_\omega
= \frac{1}{2} \epsilon_{\mu \nu\lambda\rho}F_{\mu \nu}\omega_{\lambda\rho}$, that is, writing
\begin{equation}
\label{QED-Lagrangian}
\widehat{\mathscr{L}}=\mathscr{L} + \frac{1}{2} \epsilon_{\mu \nu\lambda\rho}F_{\mu \nu}\omega_{\lambda\rho}~:=~\mathscr{L} +\Delta\mathscr{L}_\omega \,.
\end{equation}
That is because the coefficient of $\partial_0 A_i$ in the added term $\Delta\mathscr{L}_\omega$ gives the contribution $ \ast \, \omega_i$ to the conjugate momentum of $A_i$. We observe also that $\Delta\mathscr{L}_\omega$ involves $\omega_{0i}$ which are new. 

The covariant-looking $\Delta\mathscr{L}_\omega$, if it is not to affect equations of motion, has to be a total divergence. This means that
\begin{equation}
\epsilon_{\mu\nu\lambda\rho}\partial_\nu \omega_{\lambda\rho} = 0 \,,
\end{equation}
or $\omega$ as a two-form in $4$-dimensions must be closed :
\begin{equation}
d\omega=0.
\end{equation}
Hence in $\mathbb{R}^4$,
\begin{align}
\omega=d\hat{\alpha}, \qquad \hat{\alpha} : \textrm{ a one-form in $4$-dimensions}.
\end{align}

The term $\epsilon_{ijk0}F_{ij}\omega_{k0}$ in $\Delta\mathscr{L}_\omega$ may also be obtained by a Lorentz transformation of $F_{0i}$ and $\omega_{jk}$ from $\epsilon_{0ijk}F_{0i}\omega_{jk}$. The presence of $\omega_{0i}$ in $\Delta\mathscr{L}_\omega$ means that it can in general dress the electron with an infrared cloud containing both electric and magnetic fields. 

It is natural to identify the asymptotic part of $\omega_{\mu \nu}$ with Buchholz's $f_{\mu\nu}$ \cite{Buchholz2}. 

As observed in the introduction, $\Delta\mathscr{L}_\omega$ resembles the QCD $\theta$-term.

Note also that $\omega_{\mu\nu}$ depends on the charged particle. Hence it transforms under CPT in a manner required to maintain CPT invariance of \eqref{QED-Lagrangian}.   

\section{The Twisted Mass for QED$_4$}

The infrared cloud leading to Lorentz violation brightens the sphere $S^2_\infty$ at spatial infinity. We need a model for observing its effects. In setting up this model, we can be guided by the spontaneous symmetry breakdown due to the Higgs field.

In general,
\begin{equation}
  \chi(r\hat{n})\stackrel{r\to\infty}{\longrightarrow} \chi^\infty(\hat{n})=\sum\chi^\infty_{lm}~Y_{lm}(\hat{n}).
\end{equation}
Let $\hat{\chi}$ denote those test functions for which the $l=0$ term is absent,
\begin{equation}
\hat{\chi}(r\hat{n})\stackrel{r\to\infty}{\longrightarrow}\hat{\chi}^\infty(\hat{n})=\sum_{l\neq 0}\hat{\chi}^\infty_{lm} ~Y_{lm} (\hat{n}) \,.
\end{equation}

If the charged particle state is not dressed by the infra-photons, that is, on states $|\cdot\rangle_{\omega=0}$,
\begin{equation}
  Q(\chi)|\cdot\rangle_0=\chi^\infty_{00} Q_0|\cdot\rangle_0.
\end{equation}
Hence
\begin{equation}
  Q(\hat{\chi})|\cdot\rangle_0=0.
\end{equation}
It follows that twisting the mass term $m\bar{\psi}\psi$ of spin-$1/2$ particles to
\begin{equation}
 m \cos Q (\hat{\chi}) = \frac{m}{2}\left( e^{iQ(\hat{\chi})}+e^{-iQ(\hat{\chi})} \right)\bar{\psi}\psi := m(\tilde{\chi})\bar{\psi}\psi, \qquad m(0)\equiv m
\end{equation}
 does not change the physics in the sector with label $\omega=0$ where the mass term is unaffected. Furthermore $m(\hat{\chi})\bar{\psi}\psi$ is local, since $m\bar{\psi}\psi$ is local and the twist $e^{iQ(\hat{\chi})}$ commutes with local observables by gauge invariance. This property is independent of $\omega$ labeling state vectors.
 
 The Weyl relation is still valid. If 
 \begin{equation}
   U(\hat{\chi}) = e^{iQ(\hat{\chi})},
   \end{equation}
   then
   \begin{equation}
   U(\hat{\chi})V(\omega)=c(\hat{\chi}^\infty,\omega^\infty) V(\omega)U(\hat{\chi}).
 \end{equation}
Therefore with $\omega^\infty=da$,
 \begin{equation}
   \int_{S^2_\infty} da~\chi^\infty_{00}Y_{00}=0.
 \end{equation}
 
 Now,
 \begin{equation}
   _\omega\langle \cdot| \frac{m}{2}\left( e^{iQ(\hat{\chi})}+e^{-iQ(\hat{\chi})}\right)   |\cdot \rangle_\omega = \, _0 \langle \cdot| m\cos\left( \int_{S^2_\infty}\omega^\infty\hat{\chi}^\infty\right) |\cdot \rangle_0. 
 \end{equation}
 Therefore the effect of the $\omega$-twist is to replace $m$ with
 \begin{equation}
   m\cos\left( \int_{S^2_\infty}\omega^\infty\hat{\chi}^\infty\right) \,.
   \label{twistedmass-1}
 \end{equation}

In \eqref{twistedmass-1}, $\omega^\infty$ is known (see below) while $\hat{\chi}^\infty$ is associated with the charged particle. It is a new form factor for the charged particle. It is neither rotationally nor Lorentz invariant. It also depends on the sum of the momenta of all the charged particles in, say, the in- or out-state vector. We interpret $\hat{\chi}^\infty$ as a new form factor characterising the charged particle. Let an experiment measure the mass of the charged particle in the direction ${\hat x}$. Then the resultant value for the mass is given by \eqref{twistedmass-1}. 

\section{Calculation of $\omega$}
 
The vertex operator creates a coherent state of infrared photons which depends on the charged particle state $|\cdot\rangle_0$ on which it operates. This coherent state has been calculated in a form convenient for us by Fr\"ohlich et al. \cite{Frohlich1979241}. We will use their results without proof. Other important works relevant to us on this coherent state are by Kibble \cite{Kibble1, Kibble:1969ip, Kibble:1969ep, Kibble:1969kd}, Eriksson \cite{Eriksson} and Gervais and Zwanziger \cite{Gervais}.

The results of Fr\"ohlich et al. can be described as follows. In the Coulomb gauge, the free electromagnetic field has the expression
\begin{align}
A_i(\vec{x}) &=\int \frac{d^3k}{2k_0}\left ( a_i(\vec{k})e^{i \vec{k}\cdot\vec{x}} + a_i^\dagger(\vec{k}) e^{-i \vec{k}\cdot\vec{x}}\right ) \,,
\end{align}
where
\be
k_0 = |\vec{k}| \,,  \qquad A_0 = 0 \,, \qquad k_i ~ a_i(\vec{k}) = 0 \,.
\ee 
The commutation relations are
\be
\left \lbrack a_i(\vec{k}), a_j^\dagger(\vec{k}^\prime) \right \rbrack = \left(\delta_{ij}-\hat{k}_i\hat{k}_j \right)~2k_0~\delta^3(\vec{k}-\vec{k}'), \qquad \hat{k}_i =\frac{k_i}{|\vec{k}|} \,.
\ee

Let $|\vec{p},e \rangle |0 \rangle$ denote the tensor product of a single non-interacting charged particle vector of charge $e$ and momentum $\vec{p}$ and the photon vacuum, where we have suppressed spin labels. Then the dressed electron state vector surrounded by its infrared cloud is 
 \begin{equation}
 V(\omega_p)|\vec{p},e\rangle |0\rangle \,,
 \end{equation}
 where as indicated, $\omega$ depends on $\vec{p}$.
 
 Let us write 
 \begin{equation}
   \ast \omega_p(x)_i=\int d^3k \ast \widetilde{\omega}_p(k)_i e^{i\vec{k}\cdot\vec{x}}.
    \end{equation}
Then the Fourier transform $\ast {\tilde \omega}_p(k)_i$ is \cite{Frohlich1979241}
\begin{equation}
\ast \widetilde{\omega}_p(k)_i = \frac{e}{p \cdot k}\left(\vec{p}_i-\vec{p}\cdot \hat{k}~\hat{k}_i\right).
\end{equation}
Since
\begin{equation}
  V(\omega_p)^{-1}~a_i(k)~V(\omega_p) = a_i(k) + i \left (\ast \widetilde{\omega}_p(k)_i \right) \,, 
\end{equation}
we can set $V(\omega)= \{\mathds{1}\}$ and instead replace $A_i$ by $\sigma_p(A_i)$ where, as in \cite{Frohlich1979241}
    \begin{equation}
	    \label{coherent-state-map-1}
      \sigma_p(a_i)(k)=a_i(k)+ i \left (\ast \widetilde{\omega}_p(k)_i \right) \,.
    \end{equation}
The map
\begin{equation}
  \sigma_p:~a_i~\mapsto~\sigma_p(a_i)
\end{equation}
and its adjoint define an automorphism of the algebra of creation and annihilation operators.

For $N$ charged particles of momenta $\vec{p}_a$ and charges $e_a$ $(a: 1\ ,,\cdots \,, N)$, the dressed state vector is
\begin{equation}
V \left(\sum_a \omega_{\vec{p}_a} \right)\prod_i |\vec{p}_a,e_a \rangle | 0 \rangle \,, 
\end{equation}
where the Fourier transform of $\ast \omega_{p_a}$ is 
\begin{equation}
\ast \widetilde{\omega}_{p_a}(k)=\frac{e_a}{p_a \cdot k}\left(\vec{p}_a - \vec{p}_a \cdot \hat{k} ~\hat{k} \right) \,.
\end{equation}

Lorentz invariance can be spoilt by the presence of the second term in \eqref{coherent-state-map-1}. For the case of a single charged particle, for Lorentz invariance, if $U(1)$ is the unitary operator for Lorentz transformation $\Lambda$ on the quantum Hilbert space, we require that
\begin{align}
 U(\Lambda) V(\omega_p) |\vec{p}, e \rangle | 0 \rangle=V(\omega_{\Lambda p}) | \Lambda \vec{p} \,, e \rangle |0 \rangle \,,
\end{align}
However, this is not the case. Instead, we have that \cite{Frohlich197961, Frohlich1979241}
\begin{equation}
U(\Lambda) V(\omega_p) U(\Lambda)^{-1} = V(\omega_{\Lambda p}+\Omega_e) \,,
\end{equation}
where
\begin{align}
\label{Big-Omega-1-particle}
\ast \Omega_e(x)_i &=\int d^3 k~{} \ast \widetilde{\Omega}_e(k)_i e^{i\vec{k}\cdot\vec{x}}, \\
\label{Big-Omega-2-particle}
 \ast \widetilde{\Omega}_e(k)_i&=-\frac{e}{(\Lambda k)_0}\left ( (\Lambda^{-1})_{0i}-(\Lambda^{-1})_{0j}\hat{k}_j \hat{k_i} \right ) \,.
\end{align}
The unit vector $\hat{k}$ denotes the direction in which we observe the "sky" of the source. Thus Lorentz symmetry is broken. 

It is remarkable that the Lorentz invariance breaking second term in \eqref{Big-Omega-2-particle} has no dependence on charged particle momentum. As a consequence, in the $N$ charged particle sector, this term depends only on the total charge $Q$ and $\hat{k}$. Thus,
\begin{equation}
  U(\Lambda) V\left(\sum_i \omega_{p_a}\right) U(\Lambda)^{-1} = V\left(\sum_a \omega_{(\Lambda p)_a}+\Omega_Q\right),
\end{equation}
where 
\begin{equation}
  Q=\sum_a e_a
\end{equation}
and $ \ast \widetilde{\Omega}_Q(k)$ is given by \eqref{Big-Omega-2-particle} with $Q$ for $e$.

Writing
\begin{equation}
P_a = \sum_a p_a \,,
\end{equation}
as the total charged particle momentum, we can capture the features of Lorentz invariance breaking in the $N$ charged particle sector by replacing $V(\sum \omega_{p_a})$ by $V(\omega_P =\omega_{\sum p})$. As regards Lorentz invariance, we loose no information since 
\begin{equation}
  V \left ( \sum_a \omega_{p_a}-\omega_P \right ) \,,
\end{equation}
does not induce Lorentz invariance violation.

In scattering theory, however, with widely separated particles in the in or out state vector, the above replacement may not be appropriate as it is nonlocal. Instead, it seems best to use $V( \sum_a \omega_{p_a})$ which dresses each charged particle with its own infrared cloud.

If $\Lambda$ is a rotation, $ \ast \Omega_Q(k)$ vanishes. This suggests that  the rotational symmetry is preserved. However,  we note that composition of two boosts can produce a rotation, suggesting trouble with the latter too. This point requires further clarification.

\section{Consequences of Twisted Electron Mass}

The consistency of twisting the charged spinor mass to capture Lorentz breaking raises issues about its effect on locality and perturbation theory. We tentatively suggest the following answers.

As we observed, $Q(\hat{\chi})$ commutes with local observables. Hence locality seems unaffected.

In perturbation theory, we encounter the charged particle propagator
\begin{equation}
  \left\langle T\left(\psi(x)\overline{\psi}(y)\right) \right\rangle
\end{equation}
in the internal lines. This is a vacuum expectation value and the twisted mass becomes the untwisted one on vacuum. So it appears that no internal lines are affected. Hence the Lorentz breaking term affects only external lines and thereby scattering amplitudes.

This conclusion is supported by the Lagrangian \eqref{QED-Lagrangian}. The Lorentz breaking term is a surface term at infinity. 

The twisted mass term 
\begin{equation}
m \cos\left(Q(\hat{\chi})\overline{\psi}\psi \right)
\end{equation}
is a local operator. Its correlators restricted to local space-time regions should not be affected by the twist.

We can try to observe the effect of mass twist in dispersion relation and in scattering where the electron is the dressed one. Then the mass term $m$ of the electron with momentum ${\vec p}$ in the scattering amplitude is changed to
\begin{equation}
	\label{changed-mass-1}
m ({\vec{p}}\,, {\hat \chi}) \equiv m \cos\left( \lim_{r \rightarrow \infty}\int_{S^2_\infty}  r^2 d\Omega_{\hat{x}} {\hat \omega}_p^\infty (x) {\hat \chi}^\infty (\hat{x}) \right) \,.
\end{equation}
where
\be
\lim_{r \rightarrow \infty} {\hat \omega}_p^\infty (x) =  \lim_{r \rightarrow \infty} r^2 d\Omega_{\hat{x}} \ast \omega_p^\infty (x)_i  {\hat x}_i\,.
\ee
\eqref{changed-mass-1} is different for incident and outgoing momenta, if they differ. 

Here, we are twisting each charged particle mass by its own infrared cloud. As discussed earlier, we can also twist the entire $N$-particle state vector depending on the total momentum $\vec{P}$ and the total charge $Q$. Then $P_0^2-\vec{P}^2$ is the variable $s$ of scattering theory. The results below can easily be adapted to this case.

The dispersion relation reads
\be
{\vec p}^{~2} + m^2 ({\hat p}\,, {\hat \chi}) =p_0^2 \,.
\ee

Now, as noted previously, we have 
\begin{align}
\label{omega-int-5}
 \ast \omega_p(x)_i &=\int d^3k~{} \ast \widetilde{\omega}_p(k)_i e^{i\vec{k}\cdot\vec{x}}, \\
 \ast \widetilde{\omega}_p(k)_i &=\frac{e}{p \cdot k}\left(p_i - \vec{p} \cdot \hat{k} \, \hat{k}_i \right) = \frac{e}{k_0 (p_0 - \vec{p}\cdot \hat{k})} 
 \left(p_i - \vec{p} \cdot \hat{k} \, \hat{k}_i \right)\,.
\end{align}

We can take the large $r$ limit of the expression \eqref{omega-int-5} following Gervais and Zwanziger \cite{Gervais}. Thus, write 
\begin{equation}
\ast \omega_p(x)_i = \int d\Omega_{{\hat k}} \omega^2 d\omega \left( \ast \widetilde{\omega}_p(\hat{k})_i e^{i\omega r (\hat{k} \cdot \hat{x} + i\epsilon)}\right), \qquad \epsilon>0,
\end{equation}
where we use $k_0 = \omega$ and 
\begin{equation}
\ast \widetilde{\omega}_p(\hat{k})_i = \frac{e}{\omega (p_0 - \vec{p}\cdot \hat{k})} \left(p_i - \vec{p} \cdot \hat{k} \, \hat{k}_i \right) \,,
\end{equation}
and $\epsilon>0$ gives the high frequency cut-off. Hence
\begin{equation}
  \label{omega-int-6}
\ast \omega_p(x)_i=\lim_{\epsilon\to 0^+}\int d\Omega_{{\hat k}} \omega d\omega \left(\frac{e}{(p_0 - \vec{p}\cdot \hat{k})} \left(p_i - \vec{p} \cdot \hat{k} \, \hat{k}_i \right) e^{i\omega r (\hat{k}\cdot \hat{x} + i\epsilon)}\right) \,,
\end{equation}
or defining $\omega^\prime = \omega r$,
\begin{equation}
  \label{omega-int-7}
 \ast \omega_p(x)_i = - \frac{1}{r^2}\lim_{\epsilon\to 0^+}\int d\Omega_{{\hat k}} \frac{e}{(p_0 - \vec{p}\cdot \hat{k})} \left(p_i - \vec{p} \cdot \hat{k} \, \hat{k}_i \right) \left( \frac{1}{\hat{k}\cdot\hat{x}+i\epsilon}\right)^2 \,.
\end{equation}
This gives \eqref{changed-mass-1}. The pole in \eqref{omega-int-7} at $p_0 = \vec{p} \cdot \hat{k}$ can be treated as a principal value while the 
pole due to the $\left(\hat{k}\cdot\hat{x}+i\epsilon\right)^{-2}$ term gives a well-defined integral by the $\epsilon \to 0^+$ prescription.

\subsection{Spontaneous Breakdown of Symmetries: Internal and Spacetime}

In the standard model, the Higgs field breaks $SU(2) \times U(1)$ to a $U(1)$ subgroup. We can understand this result by noting that in quantum 
field theory, the group generators of  the broken transformations diverge because of the asymptotic Higgs field.

We shall now demonstrate that the Lorentz boost generators diverge in the charged sectors with infrared dressing. Hence the mechanism 
breaking Lorentz boosts is similar to spontaneous breaking of internal symmetries.

This calculation also shows that angular momenta and four-momenta do not show such a divergence.This result is consistent with those of \cite{Frohlich1979241}.

Below we give a give a quick review of spontaneous breakdown of $U(1)$ by a Higgs field using the collective coordinate approximation. We 
then examine the cases of boosts and the remaining Poincare' generators. 

We begin with some general remarks regarding spontaneous symmetry breakdown. It is followed up with the collective coordinate calculation.

Standard spontaneous symmetry breakdown is caused by the Higgs field at spatial infinity. One may imagine then that it cannot be observed by local physics. But that is not the case. It makes itself felt in two ways, under two circumstances:

\begin{itemize}

\item In a non-gauge theory, it creates Goldstone bosons. They are described by quantizing the local fluctuations around the constant Higgs field configuration. They affect the spectrum of the Hamiltonian by creating massless particles and eliminating the spectral gap of the Hamiltonian between vacuum and a particle state.
  
\item In a gauge theory, appropriate gauge fields consume the Higgs fluctuations and become massive. Thus the opposite happens regarding the Hamiltonian spectrum: a gapless spectrum with massless vector fields gets gapped. Mach suggested that inertia is affected by the ambient background \cite{Mach}. That is what happens here to the inertial mass of the vector fields. It is important to observe that the acquired mass of vector fields depend on their quantum numbers. For instance, the $W^\pm$ and $Z$ masses are different. 

\item In theories with spontaneous symmetry breaking, the passage from the massless to the massive phase of the gauge field involves the gauge transformation to the $U$-gauge. It is affected by the gauge group element obtained from the polar decomposition of the Higgs field. It preserves locality so that the massive vector theory is a local theory. We regard this also as an important fact.

\end{itemize}

Now we turn our attention to the collective coordinate calculation.

Let $\phi$ be a complex Higgs field with $\phi \rightarrow \phi_0 \neq 0$ as $|\vec{x}| \rightarrow \infty$. Let us calculate the charge operator $Q$ by 
collective coordinate method. If the classical configuration is, say
\be
\phi(\vec{x})=\phi_0\equiv \rm{constant},
\ee
we put 
\be
\phi (\vec{x},t) = e^{i \theta(t)} \phi_0
\ee
in the Lagrangian $L$. Then
\be
L=\int d^3 x (|\dot{\phi}|^2 - |\phi'|^2) = \int d^3x |i \dot{\theta} \phi_0|^2 = \dot{\theta}^2 \int d^3 x |\phi_0|^2 = \infty.
\ee
Thus the "moment of inertia" $\int d^3 x |\phi_0|^2$ diverges, showing that the symmetry $\phi \rightarrow e^{i \theta} \phi$ is spontaneously broken. 
This argument also implies that the charge
\be
Q=\int d^3 x (\pi^* \phi + \pi \phi^*)
\ee
is divergent if $\phi \rightarrow \phi_0$ as $|\vec{x}|\rightarrow \infty$ even if $\phi$ depends on $\vec{x}$.

A conceptually identical argument applies to the Lorentz boost generators, but since the details are a little different, we give them below. 

Consider the infrared-dressed state \eqref{IRdressed}
\be
V(\omega_p)|\cdot\rangle_0 = e^{i \int d^3x A_i (x) \ast \omega_p(x)_i}|\cdot\rangle_0 = |\cdot\rangle_{\omega_p}
\ee
From \eqref{omega-int-7}, we see that $\ast \omega_p(\vec{x})_i = O\left(\frac{1}{r^2}\right)$ as $r \rightarrow \infty$. 

The boosts $K_i$ are given by
\be
K_i = \int d^3 x\, x_i \,{\cal H}(x), \quad \textrm{where} \quad {\cal H}(x) = \frac{1}{2}\left( \vec{E}^2(x)+ \vec{B}^2(x) \right).
\ee
In the state $|\cdot\rangle_{\omega_p}$, the electric field $E_i$ is shifted to $\widehat{E}_i(x) = E_i(x) + \ast \, \omega_p(x)_i$ while the magnetic field $B_i$ is unchanged. Thus the boost operators become
\be
\widehat{K}_i=V^{-1}(\omega_p)K_i V(\omega_p) = \frac{1}{2}\int d^3 x \, x_i \,\left( (\vec{E}(x)+\ast \vec{\omega}_p(x))^2 + \vec{B}^2(x) \right)
\ee
when the vertex operator is absorbed in the redefined boosts $\widehat{K}_i$.

This has the divergent field-independent term 
\be
\int d^3x \,x_i (\ast \vec{\omega}_p(x))^2\,.
\ee
This diverges because $x_i \sim O(r)$ and $(\ast \vec{\omega}_p(x))^2 \sim O(1/r^4)$ as $r \rightarrow \infty$ 
(see \eqref{omega-int-7}).

The finiteness of the remaining field-dependent terms depends on the states they act on. For instance, they give finite answers on Fock state states.

On the other hand, for the transformed Hamiltonian
\be
\widehat{H} = \frac{1}{2}\int d^3x \left((\vec{E}(x)+\ast \vec{\omega}_p(x))^2+\vec{B}^2 \right) \,, 
\ee
the field-independent term gives a finite integral. The cross-term in the above integral, that is, 
\begin{equation}
\label{cross-terms}
  \int d^3 x ~\vec{E}(x) \cdot \ast\vec{\omega}_p(x)  
\end{equation}
is also well-defined for free electric field and acting on Fock space states. Indeed, recall that for a free field $\phi$, the integral $\int d^3x \phi(\vec{x})\alpha(\vec{x})$ acting on the vacuum gives a square-integrable state if $\alpha$ is square-integrable. This operator is similarly checked to be well-defined on any Fock space state. In our case $\ast \vec{\omega}$ plays the role of $\alpha$ is square-integrable. We conclude that $\widehat{H}$ is a well-defined operator. Actually, it is unitarily equivalent to $H$.

The fact that the energy stored in the infrared cloud is finite is well-known and explained in QFT textbooks as Itzykson and Zuber \cite{Itzykson-book} (Section 4.1.2) and Peskin and Schroeder \cite{Peskin-book} (Section 6.1).

Next consider the angular momenta
\begin{equation}
  J_i=\int d^3x~ E_j \left( -i \delta_{jk}\partial_i - i \epsilon_{ijk} \right) A_k.
\end{equation}
It is transformed as
\begin{equation}
  J_i \rightarrow \widehat{J}_i = V^{-1}(\omega_p)J_i V(\omega_p)
\end{equation}
by $V(\omega_p)$ with no field-dependent term. The same is true for the momenta 
\begin{equation}
  P_i =\int d^3x ~E_j \left(  -i\partial_i \right) A_j .
\end{equation}
So we can show as in the case of \eqref{cross-terms} that these operators are well-defined in the Fock space.

Finally we infer that only boosts are spontaneously broken. This is compatible with \cite{Frohlich1979241}.

\subsection{On Lorentz-Violating Mass: Dispersion Relation and Particle Spin are Changed}

We can write \eqref{changed-mass-1} as follows
\be
m ({\vec{p}}\,, {\hat \chi}) \equiv m \cos \left ( e \int d\Omega_{\hat{x}} \int d\Omega_{\hat{k}}  {\hat \chi}^\infty (\hat{x})   
\frac{{\vec{p}}\cdot {\hat x} - \vec{p} \cdot \hat{k} \, \hat{k} \cdot {\hat x}}{(p_0 -\vec{p} \cdot \hat{k}) \, (\hat{k}\cdot\hat{x} + i \epsilon)^2} \right) \,.
\label{twistedmass}
\ee
We would like to remark that the photon cloud surrounding  the charged particle has the electric field $\ast \omega_p(x)_i$ as alluded to after \eqref{twistedmass-1}. The function ${\hat \chi}^\infty$ takes moments of this field: they are determined by the experimental arrangement measuring say the mass. 

The physical interpretation of ${\hat \chi}^\infty$ is that it is a new form factor of the charged particle.

Let $R \in SO(3)$. It acts on the function $\hat{\chi}$ as  $(R{\hat \chi})(x)={{\hat \chi}(R^{-1} x)}$. Then by using the rotational invariance of the two measures, we get
\be 
m( R{\vec{p}} \,, R {\hat \chi}) =m({\vec{p}} \,, {\hat \chi}) \,.
\label{rotinv}
\ee
Thus $m$ is a rotationally invariant function of its arguments.

From each ${\hat \chi}_{lm}$, we can form a scalar by coupling it to a polynomial of degree $l$ in $\vec{p}$. The integral in \eqref{twistedmass} is a linear combination of these scalars. The simplest $\hat{\chi}^\infty$ to consider has only $l=1$, its $l=0$ component being 0. Retaining just $l=1$, we henceforth set 
\be
{\hat \chi}^\infty (\hat{x}) = \chi_{i} {\hat x}_i, \quad \chi_i = \textrm{real constants}. 
\ee
Then the integral takes the form
\be   
C(\vec{p}^2) \chi_{i} {p}_i
\label{I1result}
\ee
for $\epsilon \rightarrow 0^+$. We evaluate $C(\vec{p}^2)$ in the Appendix. It is finite as $\vec{p} \rightarrow 0$:
\begin{equation}
C(\vec{p}^2)= 8\pi^2 e \frac{p_0^2-|\vec{p}|^2}{|\vec{p}|^3} \ln \left(\frac{p_0+|\vec{p}|}{p_0-|\vec{p}|}\right) -
16\pi^2 e \frac{p_0}{|\vec{p}|^2}.
\end{equation}
Thus finally for $\epsilon \rightarrow 0^+$,
\be
m({\hat p} \,, {\hat \chi}) = m \cos (C(\vec{p}^2) \chi_{i} {p}_i)
\label{IRmass}
\ee
For normalised $\chi_{i}$, that is, for ${\tilde \chi}_{i}= \frac{\chi_{i}}{\sqrt{\chi_{k} \chi_k}}$, we can if desired plot the twisted mass say for ${\hat p} = (0,0,1)$. 

A generic Lorentz boost of ${\hat p}$ brings in more complicated functions of $p_i$. For example, \eqref{I1result} is changed to
 $I^\prime = C((\overrightarrow{\Lambda \hat{p}})^2) \chi_{i} (\Lambda \vec{p})_i$.

\subsection{Mass Twist Smears Mass and Spin}

In the Poincar\'e representation theory for a massive particle, mass is assumed to be a scalar and spin is introduced by attaching an irreducible representation (IRR) of SU(2) to the vector state in the rest frame. The twisted mass, however, depends on $p$ or $P$ and is not a rotational scalar. Its value depends on the state vector it is associated with. Thus mass gets smeared, depending on $p$ or $P$. Such a smearing {\it may} be compatible with the results of Buchholz \cite{Buchholz2}.

Further, as mentioned, the twisted mass is not a rotational scalar. By \eqref{IRmass}, it depends on $\chi_{i} \vec{P}_i$ and all its powers for the choice made above for $\chi$. Thus standard spin such as $1/2$ of the muon acquires all its orbital excitations, which depend on ${\hat \chi}^\infty$. Its $(2n)^{th}$ power is suppressed by the coefficient $\alpha^n$, with $\alpha$ being the fine structure constant. This phenomenon will affect decay selection rules (and of courses scattering). Further analysis of this observation is called for. 

For sensitive experiments on the isotropy of space, see \cite{Muller:2007zz, Herrmann:2005qe, Brillet}.

\section{Non-Abelian Superselection and Higgs Symmetry breaking}

Non-Abelian super-selection rules play a role even in the familiar phenomenon where a complex Higgs field breaks a $U(1)$ symmetry spontaneously. We conclude this paper with this observation.

We consider $U(1)$ gauge symmetry broken by a complex scalar field $\varphi(x)$. Let $f_R$ be test functions supported in $r \geq R$. Then 
\be
S_R = \int d^3 x {\bar f}_R(x) \varphi(x) \,,
\label{sr}
\ee
commutes with all observables supported in $r < R$. So $S_R$ for $R \rightarrow \infty$, denoted by $S_\infty$, is superselected. We note that
\be
Q(\xi) = \int d^3 x ( \partial_i \xi E_i + \xi J_0) \,,
\label{qxi}
\ee
is superselected too and \eqref{sr} and \eqref{qxi} do not commute. We have that
\be
e^{i Q (\xi)} S_\infty e^{i Q (\xi)} = e^{i \xi_\infty} S_\infty \,, \quad \xi_\infty = \xi |_\infty \,.
\ee

We assume that $S_\infty \neq 0$. 

Both $S_\infty$ and $Q(\xi)$ commute with all local observables. Therefore, our previous arguments \cite{Bal-Sachin}   lead to the conclusion that one of them must be spontaneously broken.

We explain the above argument briefly in our context. If we diagonalize $S_\infty$, that defines a superselection sector. However, $Q(\xi)$ then changes it by \eqref{qxi}. Hence it is spontaneously broken.

If we diagonalize $S_\infty$, as we do in superconductivity, we get a domain ${\cal D}_1$ for the Hamiltonian ${\cal H}$ which makes it self-adjoint. The operator $e^{i Q (\xi)}$ changes this domain: it is spontaneously broken.

We can also opt to diagonalize $e^{i Q (\xi)}$. Then $U(1)$ is preserved, but at the expense that ${\cal H}$ is no longer defined: the integral of energy density diverges classically. We can still define $e^{-i t{\cal H}}$, finding first a domain in which $S_\infty$ is diagonal and then extending this unitary operator to the vectors with $e^{i Q(\xi)}$ diagonal. (A unitary operator can act on all vectors of the Hilbert space.) We call such a domain ${\cal D}_2$.

If ${\cal P}_E$ is the projector for energies less than $E$ for ${\cal H}$ on ${\cal D}_1$, it is bounded and hence defined in ${\cal D}_2$. A legitimate question is, whether if we reconstruct ${\cal H}$ from ${\cal D}_2$, then we will get the same or similar low energy spectrum of ${\cal H}$ as from ${\cal D}_1$. If that is the case, then there may be a new approach to spontaneous symmetry breaking.

\section*{Acknowledgments}

We acknowledge discussions on possible experimental realization of our results, specially muon decay as a function of direction towards sky with Catalina Curceanu (INFN, Frascati), Edoardo Milotti (Universit\`a di Trieste) and Sudhor Vempati (CHEP, IISc, Bangalore). A.P.B. thanks T\"UBiTAK and Seckin K\"{u}rk\c{c}\"{u}o\v{g}lu for hospitality at the Middle East Technical University,Ankara. He also thanks the group at the Centre for High Energy physics, Indian Institute of Science, Bangalore, and especially Sachin Vaidya, for hospitality. S.K. thanks A. Behtash for useful discussions. S.K. is supported by T\"UBA-GEBiP program of the Turkish Academy of Sciences. A.R.Q. is supported by CAPES process number BEX 8713/13-8. 

\appendix

\section{Calculation of the integral in \eqref{twistedmass}}
We wish to evaluate
\be
I=e \int d\Omega_{\hat{x}} \int d\Omega_{\hat{k}}  {\hat \chi}^\infty (\hat{x})   
\frac{{\vec{p}}\cdot {\hat x} - \vec{p} \cdot \hat{k} \, \hat{k} \cdot {\hat x}}{(p_0 -\vec{p} \cdot \hat{k}) \, (\hat{k}\cdot\hat{x} + i \epsilon)^2} \, .
\ee
For $\hat{\chi}^\infty(x) = \chi_i \hat{x}_i$, $I$ is $C(\vec{p}^2) \chi_i p_i$. To evaluate $C(\vec{p}^2)$, we choose $\chi_i=p_i$ to find
\beqa
C(\vec{p}^2) \vec{p}^2 &=& e \int d\Omega_{\hat{x}} \int d\Omega_{\hat{k}} (\vec{p} \cdot \hat{x})
\frac{{\vec{p}}\cdot {\hat x} - \vec{p} \cdot \hat{k} \, \hat{k} \cdot {\hat x}}{(p_0 -\vec{p} \cdot \hat{k}) \, (\hat{k}\cdot\hat{x} + i \epsilon)^2}, \nn \\
&=& e \int d\Omega_{\hat{k}} \frac{1}{(p_0 -\vec{p} \cdot \hat{k})} \int d\Omega_{\hat{x}} \frac{({\vec{p}}\cdot {\hat x})^2}{(\hat{k}\cdot\hat{x} + i \epsilon)^2} \nn \\
&& \quad \quad \quad  - e\int d\Omega_{\hat{k}} \frac{\vec{p} \cdot \hat{k} }{(p_0 -\vec{p} \cdot \hat{k})} \int d\Omega_{\hat{x}} \frac{\vec{p} \cdot \hat{x} \, \hat{k} \cdot {\hat x}}{(\hat{k}\cdot\hat{x} + i \epsilon)^2}, \nn \\
&\equiv& I_1-I_2 
\eeqa
The integral $I_1$ is
\be
I_1=e \int d\Omega_{\hat{k}} \frac{p_i p_j}{(p_0 -\vec{p} \cdot \hat{k})} \int d\Omega_{\hat{x}} \frac{\hat{x}_i \hat{x}_j}{(\hat{k}\cdot\hat{x} + i \epsilon)^2}
=e \int d\Omega_{\hat{k}} \frac{1}{(p_0 -\vec{p} \cdot \hat{k})} K_1
\ee
where 
\beqa
K_1 &=& p_i p_j \int d\Omega_{\hat{x}} \frac{\hat{x}_i \hat{x}_j}{(\hat{k}\cdot\hat{x} + i \epsilon)^2} = p_i p_j (\alpha \delta_{ij} + \beta \hat{k}_i \hat{k}_j), \\
\alpha \delta_{ij} + \beta \hat{k}_i \hat{k}_j &=& \int d\Omega_{\hat{x}} \frac{\hat{x}_i \hat{x}_j}{(\hat{k}\cdot\hat{x} + i \epsilon)^2} \label{alphabeta}.
\eeqa
Here, $\alpha$ and $\beta$ are rotational scalars by rotational invariance. To find $\alpha$ and $\beta$, multiply LHS of \eqref{alphabeta} by 
$\delta_{ij}$ and sum over $i,j$ to get
\be
3\alpha+\beta = \lim_{\epsilon \rightarrow 0} \int d\Omega_{\hat{x}} \frac{1}{(\hat{k}\cdot\hat{x} + i \epsilon)^2} =- \lim_{\epsilon \rightarrow 0} \frac{4\pi}{1+\epsilon^2} = -4\pi
\ee
Next, multiply LHS of \eqref{alphabeta} by $\hat{k}_i \hat{k}_j$ and sum over $i,j$ to get
\be
\alpha +\beta =  \lim_{\epsilon \rightarrow 0} \int d\Omega_{\hat{x}} \frac{(\hat{k} \cdot \hat{x})^2}{(\hat{k}\cdot\hat{x} + i \epsilon)^2} = 4\pi.
\ee
Solving for $\alpha$ and $\beta$, we get
\be
\alpha =-4\pi, \quad \beta=8\pi,
\ee
and
\be
I_1=e \int d\Omega_{\hat{k}} \frac{-4\pi \vec{p}^2+8\pi(\vec{p}\cdot\hat{k})^2}{(p_0 -\vec{p} \cdot \hat{k})}
\ee
This can be evaluated by elementary methods to give
\be
I_1= 8\pi^2 e \left( \frac{2p_0^2}{|\vec{p}|} - |\vec{p}| \right)  \ln \left( \frac{p_0 +|\vec{p}|}{p_0-|\vec{p}|} \right)  - 32\pi^2 e p_0
\ee

The integral $I_2$ is
\be
I_2 = e\int d\Omega_{\hat{k}} \frac{(\vec{p} \cdot \hat{k})p_i \hat{k}_j}{(p_0 -\vec{p} \cdot \hat{k})} \int d\Omega_{\hat{x}} \frac{\hat{x}_i \, {\hat x}_j}{(\hat{k}\cdot\hat{x} + i \epsilon)^2}
\ee
The angular integral over $\hat{x}$ can be done as for \eqref{alphabeta}. One finds that it is $-4\pi \delta_{ij}+8\pi \hat{k}_i \hat{k}_j$, giving
\be
I_2 = 4\pi e\int d\Omega_{\hat{k}}\frac{(\vec{p} \cdot \hat{k})^2}{(p_0 -\vec{p} \cdot \hat{k})} = 8\pi^2 e\left(\frac{p_0^2}{|\vec{p}|} \ln \left(\frac{p_0+|\vec{p}|}{p_0-|\vec{p}|}\right) -2p_0 \right).
\ee
Finally,
\be
C(|\vec{p}|^2)=\frac{I_1-I_2}{|\vec{p}|^2}= 8\pi^2 e \frac{p_0^2-|\vec{p}|^2}{|\vec{p}|^3} \ln \left(\frac{p_0+|\vec{p}|}{p_0-|\vec{p}|}\right) -
16\pi^2 e \frac{p_0}{|\vec{p}|^2}.
\ee
It is easy to see that 
\be
\lim_{|\vec{p}| \rightarrow 0} C(|\vec{p}|^2) = -\frac{32\pi^2}{3p_0}.
\ee

\end{document}